\documentclass[12pt]{article}
\usepackage{graphicx}

\begin{document}

\title{\bf Hidden antiunitary symmetry behind ``accidental'' degeneracy and its protection of degeneracy}
\author{\small Jing-Min Hou$^{1,*}$\ \  and\ Wei Chen$^2$\\
{\it\small $^1$ School of Physics, Southeast University, Nanjing
211189, China}\\
{\it\small $^2$ College of Science, Nanjing University of Aeronautics and Astronautics,}\\{\it \small Nanjing 210016, China}\\
{\it \small Corresponding author. E-mail:$^*$ jmhou@seu.edu.cn}}

\date{}

\maketitle

\begin{abstract}
 In quantum mechanics, accidental degeneracy refers to energy
degeneracy that occurs coincidentally, without any protection by
symmetry. Here, we prove a theorem stating that any two-fold degeneracy
(accidental or not) in a quantum system is protected by a novel
hidden symmetry, which can be expressed by an antiunitary operator
with its square being $-1$. In this sense, the so-called accidental
degeneracy is not really accidental, and this actually implies a
hidden antiunitary symmetry.
\end{abstract}



 \section{Introduction}

 Energy degeneracy in a quantum system is
 generally protected by a symmetry, such as point group or time-reversal
 symmetry. Nevertheless, it is believed that there exists a type of
 degeneracy called accidental degeneracy that does not need the
 protection of symmetry. Accidental degeneracy was discussed in
 the early days of
 quantum theory by von Neumann and Wigner\cite{vNW}. To ensure that two eigenenergies of a
 $2\times 2$ Hamiltonian matrix are identical, three independent
 equations should be satisfied. Thus, three real tunable parameters can guarantee the
 existence of a two-fold degeneracy. This type of
 degeneracy is usually considered to be accidental. Later, Herring applied this theory to the
energy bands of crystals and developed the theory of accidental
band degeneracy\cite{Herring}.

Some degeneracies in quantum systems that were previously considered
to be accidental were later found to be protected by a higher hidden
symmetry\cite{Mcintosh}. For example, the accidental degeneracy in
the Kepler problem\cite{Fock}, an isotropic harmonic
oscillator\cite{Alliluev}, a free particle enclosed by an
impenetrable cubic
 box\cite{Leyvraz,Lemus}, and a quantum mechanical rigid rotor\cite{Smith} can be explained by expanding the symmetry
 group. However, not all the accidental degeneracies in quantum
 systems can be explained by a larger symmetry group. Whether a hidden symmetry exists for any accidental degeneracy is still an open question. Thus, a
 general explanation of all the accidental degeneracies is
desirable.

In this paper, we provide insight into the interpretation of accidental
degeneracy in quantum systems. First, we prove a theorem stating that any
two-fold degeneracy must be protected by a hidden antiunitary
symmetry (HAS) having an operator with its square being $-1$. We
emphasize that this HAS having an operator with its square being $-1$ is
necessary and sufficient for the protection of any two-fold
degeneracy. From the theorem, we can infer that the so-called
``accidental'' degeneracy is not really accidental. In our
previous works, we found some examples in two-dimensional
lattices, where the HAS is responsible for the degeneracy at the
Dirac points\cite{Hou1,Hou2}.

Because our theory of HAS protection of energy degeneracy is
applicable to any type of two-fold degeneracy, it is important to
clarify its relationship and compatibility with other theories.
 Most non-accidental two-fold degeneracies
are believed to be protected by the conventional symmetry group. For these
degeneracies, we show that the HAS protection is the minimum
constraint for the existence of a two-fold degeneracy and that the
constraints of the conventional symmetry group may be redundant.

 \section{Theorem}\label{theorem}
First, we prove the following theorem:
   \em
In a quantum system, any two-fold degeneracy must be protected by a
symmetry that has an antiunitary operator with its square being $-1$. \em

\emph{Proof}--We first suppose that $H$ is the Hamiltonian of a
system and has two distinct eigenstates $|\psi_1\rangle$ and
$|\psi_2\rangle$ with the same eigenenergy $E$, i.e.,
\begin{eqnarray}
H|\psi_i\rangle=E|\psi_i\rangle,\label{eigen}
\end{eqnarray}
where $i=1,2$.
 That is, the system has a
two-fold degeneracy. We can always construct an operator as
\begin{eqnarray}
\Upsilon=(|\psi_2\rangle\langle\psi_1^*|-|\psi_1\rangle\langle\psi_2^*|)K,
\label{op}
\end{eqnarray}
where the star represents complex conjugation, and $K$ is the
complex conjugation operator. Obviously, $\Upsilon$ is an
antiunitary operator, and it is easy to verify that the square of the
operator $\Upsilon$ is $\Upsilon^2=-1$. We can verify the following
equations:
\begin{eqnarray}
&&\Upsilon|\psi_1\rangle=|\psi_2\rangle, \label{ups1}\\
&&\Upsilon|\psi_2\rangle=-|\psi_1\rangle.\label{ups2}
\end{eqnarray}
From Eqs. (\ref{eigen}), (\ref{ups1}), and (\ref{ups2}), we have
\begin{eqnarray}
(\Upsilon H-H\Upsilon)|\psi_i\rangle= 0.
\end{eqnarray}
  Then, we obtain $[\Upsilon, H]=0$; namely, the system
is invariant under the action of the operator $\Upsilon$.

On the other hand, we show that if $H $ is invariant under the
the action of the antiunitary operator $\Upsilon$ with
$\Upsilon^2=-1$, the system must have a two-fold degeneracy. We
assume that $|\psi_1\rangle$ is the eigenstate of the Hamiltonian $H
$ and define
\begin{eqnarray}
|\psi_2\rangle=\Upsilon|\psi_1\rangle;\label{psi2}
\end{eqnarray}
 then
$|\psi_1\rangle$ and $|\psi_2\rangle$ have the same eigenenergy and
  may be the same eigenstate up to a factor or orthogonal to each
other. Because $\Upsilon^2=-1$ and Eq. (\ref{psi2}) are satisfied, we
have
\begin{eqnarray}
\Upsilon|\psi_2\rangle=-|\psi_1\rangle.\label{psi1}
\end{eqnarray}
 Suppose
$|\tilde\psi_1\rangle=\Upsilon|\psi_1\rangle$ and
$|\tilde\psi_2\rangle=\Upsilon|\psi_2\rangle$; then, for the
antiunitary operator, we have
\begin{eqnarray}
\langle
\tilde\psi_1|\tilde\psi_2\rangle=\langle\psi_1|\psi_2\rangle^*=\langle\psi_2|\psi_1\rangle.\label{at1}
\end{eqnarray}
Furthermore, from Eqs. (\ref{psi2}) and (\ref{psi1}), we obtain
\begin{eqnarray}
\langle
\tilde\psi_1|\tilde\psi_2\rangle=-\langle\psi_2|\psi_1\rangle.\label{at2}
\end{eqnarray}
Comparing Eq. (\ref{at2}) with Eq. (\ref{at1}), we arrive at
$\langle \psi_2|\psi_1\rangle=0$; i.e., the two eigenstates are
orthogonal to each other. Considering that they have the same
eigenenergy, we conclude that there must exist a two-fold
degeneracy.

 On the basis of the above theorem, we infer that real accidental
degeneracy does not exist in any system. The degeneracy is
considered to be accidental only because the symmetry behind it, which indeed exists, has not
been found. The so-called
 ``accidental'' degeneracy is generally obtained by fine-tuning the parameters to
satisfy some conditions, which means that the parameters obey some
constraints. Actually, these constraints are equivalent to a HAS
invariance of the Hamiltonian, which will be discussed in detail in
the next section.

\section{ Example: a two-level system    }

Now, we turn to a two-level quantum system with the Hamiltonian
$H(\alpha)$, where $\alpha$ represents all of the tunable parameters.
Without loss of generality, the Hamiltonian of the quantum system
can always be written in the form
\begin{eqnarray}
H( {\alpha})&=&h_0( {\alpha})I_2+h_x( {\alpha})\sigma_x+h_y( {\alpha})\sigma_y+h_z( {\alpha})\sigma_z\nonumber\\
&=&\left(\matrix{h_0( {\alpha})+h_z( {\alpha})&h_x( {\alpha})-ih_y(
{\alpha})\cr h_x(\mathbf{\alpha})+ih_y(
{\alpha})&h_0(\mathbf{\alpha})-h_z( {\alpha})}\right),
\label{Halpha}
\end{eqnarray}
where $\sigma_{x,y,z}$ are the Pauli matrices, and $I_2$ is the
$2\times 2$ unit matrix; $h_i(\mathbf{\alpha}), (i=0,x,y,z)$ are
real functions of $\mathbf{\alpha}$. The eigenenergies are
$E_\pm(\mathbf{\alpha})=h_0(\mathbf{\alpha})\pm\sqrt{h_x(\mathbf{\alpha})^2+h_y(\mathbf{\alpha})^2+h_z(\mathbf{\alpha})^2}$.
By fine-tuning the parameters to intersect the energy levels,
one may obtain an energy degeneracy for the parameter
$\alpha=\alpha_0$, which is generally considered to be accidental. From the expression of the eigenenergies, we can conclude
that if energy degeneracy occurs for $\alpha=\alpha_0$, $h_x,h_y$ and $h_z$ must be constrained as follows:
\begin{eqnarray}
&&h_x(\mathbf{\alpha}_0)=0,\label{cx}\\
&&h_y(\mathbf{\alpha}_0)=0,\label{cy}\\
&&h_z(\mathbf{\alpha}_0)=0.\label{cz}
\end{eqnarray}
In the following, we will show that these constraints on the
Hamiltonian are equivalent to a HAS having an operator
with its square being $-1$.

  We assume
that $\Upsilon$ is an antiunitary operator and $\Upsilon^2=-1$. We
suppose that when the parameter $\alpha=\alpha_0$, the Hamiltonian
$H(\alpha_0)$ is invariant under the action of $\Upsilon$; i.e.,
$\Upsilon H(\alpha_0)\Upsilon^{-1}=H(\alpha_0)$. The representation
of $\Upsilon$ can be written as
\begin{eqnarray}
\Upsilon=\left(\matrix{0&-1\cr 1&0}\right)K.
\end{eqnarray}
Then, under the action of the operator $\Upsilon$, we have
\begin{eqnarray}
&&\Upsilon H(\alpha_0)\Upsilon^{-1}=
\left(\matrix{h_0(\alpha_0)-h_z(\alpha_0)&-h_x(\alpha_0)+ih_y(\alpha_0)\cr
-h_x(\alpha_0)-ih_y(\alpha_0)&h_0(\alpha_0)+h_z(\alpha_0)}\right).
\label{Htrans}
\end{eqnarray}
Comparing Eqs. (\ref{Halpha}) and (\ref{Htrans}) in terms of the
identity $\Upsilon H(\alpha_0)\Upsilon^{-1}=H(\alpha_0)$, we obtain
$h_x(\alpha_0)=h_y(\alpha_0)=h_z(\alpha_0)=0$, which is just Eqs.
(\ref{cx}), (\ref{cy}), and (\ref{cz}).

From the above analysis, we find that the constraints on the energy functionals for
degeneracy imply a HAS having an operator with its square being $-1$,
which can protect the energy degeneracy.

\section{Relationship between the HAS protection and a group symmetry
protection}\label{group}

In the above theorem, we showed that any two-fold degeneracy
must be protected by a HAS. If there is a two-fold
degeneracy protected by another symmetry, for example, a point group, one may then ask what the
relationship is between the point group and the HAS. In the following, we
will answer this question in detail.

If a two-fold degeneracy is protected by a group, then this group
must have at least a two-dimensional irreducible representation,
which requires a non-abelian group. We suppose that $A$ and $B$ are
representation matrices of a two-dimensional irreducible unitary
representation of the non-abelian group and do not commute with
each other; i.e., $[A, B]\neq 0$. The unitary representation
matrices can always be written as
\begin{eqnarray}
A&=&\left(\matrix{a_1+ia_2&a_3+ia_4\cr -a_3+ia_4&a_1-ia_2}\right),\label{ma}\\
B&=&\left(\matrix{b_1+ib_2&b_3+ib_4\cr
-b_3+ib_4&b_1-ib_2}\right),\label{mb}
\end{eqnarray}
where $a_i$ and $b_i$ $(i=1,2,3,4)$ are real numbers and satisfy
$\sum_{i=1}^4 |a_i|^2=1$ and $\sum_{i=1}^4 |b_i|^2=1$, respectively. The
commutation of the matrices $A$ and $B$ reads
\begin{eqnarray}
&&[A,B]=2\left(\matrix{i\lambda_1&-\lambda_2+i\lambda_3\cr
\lambda_2+i\lambda_3&-i\lambda_1}\right),\label{comAB}
\end{eqnarray}
where $\lambda_1=a_3b_4-a_4b_3$, $\lambda_2=a_2b_4-a_4b_2$, and
$\lambda_3=a_2b_3-a_3b_2$. Because the matrices $A$ and $B$ do not
commute with each other, at least one of $\lambda_i (i=1,2,3)$ is
nonvanishing.
  Because the system is invariant under transformation of the
non-abelian group, we have
\begin{eqnarray}
&&[A, H(\alpha_0)]=0,\label{comA0}\\
&&[B, H(\alpha_0)]=0.\label{comB0}
\end{eqnarray}
On the other hand, from Eqs. (\ref{Halpha}) and (\ref{ma}), we have
\begin{eqnarray}
 &&[A,
H(\alpha_0)]=
2\left(\matrix{W_1(\alpha_0)&W_2(\alpha_0)+iW_3(\alpha_0)\cr
W_2(\alpha_0)-iW_3(\alpha_0)&-W_1(\alpha_0)}\right), \label{comA}
\end{eqnarray}
where $W_1(\alpha_0)=a_3h_x(\alpha_0)-a_4h_y(\alpha_0)$,
$W_2(\alpha_0)=a_2h_y(\alpha_0)-a_3h_z(\alpha_0)$, and
$W_3(\alpha_0)=2a_2h_x(\alpha_0)-2a_4h_z(\alpha_0)$. Comparing Eq.
(\ref{comA}) with Eq. (\ref{comA0}), we obtain the following three equations:
\begin{eqnarray}
&& a_3h_x(\alpha_0)-a_4h_y(\alpha_0)=0, \label{eqa1}\\
&& a_2h_y(\alpha_0)-a_3h_z(\alpha_0)=0,\label{eqa2}\\
&& a_2h_x(\alpha_0)-a_4h_z(\alpha_0)=0.\label{eqa3}
\end{eqnarray}
Furthermore, from Eqs. (\ref{Halpha}) and (\ref{mb}), we have
\begin{eqnarray}
&&[B,H(\alpha_0)]= 2\left(\matrix{
V_1(\alpha_0)&V_2(\alpha_0)+iV_3(\alpha_0)\cr V_2(\alpha_0)-i
V_3(\alpha_0)&- V_1(\alpha_0)}\right),\label{comB}
\end{eqnarray}
where $V_1(\alpha_0)= b_3h_x(\alpha_0)- b_4h_y(\alpha_0)$,
$V_2(\alpha_0)= b_2h_y(\alpha_0)- b_3h_z(\alpha_0)$, and
$V_3(\alpha_0)= b_2h_x(\alpha_0)- b_4h_z(\alpha_0)$. Comparing Eq.
(\ref{comB}) with Eq. (\ref{comB0}), we obtain another three
  equations:
\begin{eqnarray}
&&b_3h_x(\alpha_0)-b_4h_y(\alpha_0)=0,\label{eqb1}\\
&&b_2h_y(\alpha_0)-b_3h_z(\alpha_0)=0,\label{eqb2}\\
&&b_2h_x(\alpha_0)-b_4h_z(\alpha_0)=0.\label{eqb3}
\end{eqnarray}
For $h_x, h_y, h_z$, there are six equations, Eqs.(\ref{eqa1})--(\ref{eqa3}) and (\ref{eqb1})--(\ref{eqb3}). If they have nonzero solutions, there must be only
three independent equations, and the following constraints are required:
\begin{eqnarray}
&&\lambda_1=a_3b_4-a_4b_3=0,\\
&&\lambda_2=a_2b_4-a_4b_2=0,\\
&&\lambda_3=a_2b_3-a_3b_2=0.
\end{eqnarray}
However, because the matrices $A$ and $B$ do not commute with each other,
from Eq. (\ref{comAB}), we know that at least one of $\lambda_i
(i=1,2,3)$ is nonvanishing. Thus, the only solution of $h_x, h_y, h_z$ is
\begin{eqnarray}
&&h_x(\alpha_0)=0,\\
&&h_y(\alpha_0)=0,\\
&&h_z(\alpha_0)=0,
\end{eqnarray}
which are just the necessary constraints (\ref{cx})--(\ref{cz}) on $h_x(\alpha)$, $h_y(\alpha )$, and $h_z(\alpha )$
for the occurrence of degeneracy at $\alpha=\alpha_0$.

 From the above analysis, we conclude that the minimum requirement for a two-fold degeneracy is the
non-commutation of the two representation matrices of the group, which
is equivalent to the constraint of an antiunitary operator with its square being $-1$. If the irreducible representation has more
than two representation matrices that do not commute with each
other, the constraints of the non-abelian group for the two-fold
degeneracy are redundant. Thus, the protection of the other group
symmetry does not contradict the protection of the HAS.

\section{Conclusion}
\label{con}
 In summary, we proved a theorem stating that any two-fold
degeneracy in a quantum system must be protected by an antiunitary
operator with its square being $-1$. As a result, we can infer
that there is no accidental degeneracy in practice, which challenges
conventional knowledge. To illustrate our
conclusion, we provided an example of a two-level system. Our theory provides insight into the so-called accidental
degeneracy. Furthermore, our theorem applies to not only accidental
degeneracy but also non-accidental degeneracy. We point out that, although our theorem is
about two-fold degeneracy, it is straightforward to generalize it to
$n$-fold degeneracy by defining $n-1$ independent antiunitary
operators as in Eq. (\ref{op}).

\section*{Acknowledgments}
  This work was supported
by the National Natural Science Foundation of China under Grant Nos.
11274061 and 11504171; the Natural Science Foundation of Jiangsu Province, China under Grant
No. BK20150734.

\end{document}